\documentclass[a4paper,12pt]{elsarticle}

\usepackage{hyperref}
\usepackage{caption}
\usepackage{subcaption}
\journal{Nuclear Inst. and Meth. in Phys. Res., A, NIMA-D-14-00850}

\begin{document}

\begin{frontmatter}

\title{The GALATEA Test-Facility for High Purity Germanium Detectors}


\author[a]{I. Abt}
\author[a]{A. Caldwell}
\author[a]{B. Doenmez}
\author[a]{L. Garbini}
\author[a]{S. Irlbeck}
\author[a]{B. Majorovits}
\author[a]{M. ~Palermo\corref{cor1}}
\ead{palermo@mpp.mpg.de}
\author[a]{O. Schulz}
\author[a]{H. Seitz}
%\author[a,b]{F. Stelzer}
\author[a]{F. ~Stelzer\corref{deg1}}

\address[a]{Max-Planck-Institut f\"{u}r Physik, Munich, Germany}
%\address[b]{now at Max-Planck-Institut f\"{u}r Plasmaphysik, Garching, Germany}
\cortext[deg1]{now at Max-Planck-Institut f\"{u}r Plasmaphysik, Garching, Germany}
\cortext[cor1]{Corresponding Author. Tel: +49-89-32354371}

\begin{abstract}
GALATEA is a test facility designed to investigate bulk and surface effects in high purity germanium detectors. 
A vacuum tank houses a cold volume with the detector inside. A system of three precision motorized stages allows an almost complete scan of the detector.
The main feature of GALATEA is that there is no material between source and detector.
This allows the usage of alpha and beta sources to study surface effects. 
A 19-fold segmented true-coaxial germanium detector was used for commissioning.
A first analysis of data obtained with an alpha source is presented here.
\end{abstract}


\end{frontmatter}

%\linenumbers

\section{Introduction}
\label{intro}
  High Purity Germanium (HPGe) detectors are used in many low background experiments, especially in neutrinoless double beta decay \cite{gerda} \cite{majorana} and direct dark matter~\cite{cdex}-\cite{malbek} searches. To enhance the sensitivity of such searches, future large-scale HPGe detector based experiments are foreseen \cite{haxton}. One of the limiting sources of background in such experiments is surface contamination \cite{gerda-bkg}. This requires complete understanding and characterization of the detector response to surface events. It is, therefore, desirable to completely scan the detector surfaces. The vacuum test-facility GALATEA presented here was designed to study the properties of HPGe detectors in detail, with a focus on surface effects \cite{surface}-\cite{simpson}. In order to perform these studies, minimally penetrating radioactive sources are used to scan the detector surfaces. The sources are collimated towards the detector. There is no material on the path of the probing particles, which is a novel feature. 
%A good understanding of the detector response is, in general, essential for further detector development \cite{agostini}. 

\section{The experimental setup}
\label{setup}
\begin{figure}[!ht]
  \begin{center}
    \includegraphics[scale=0.3]{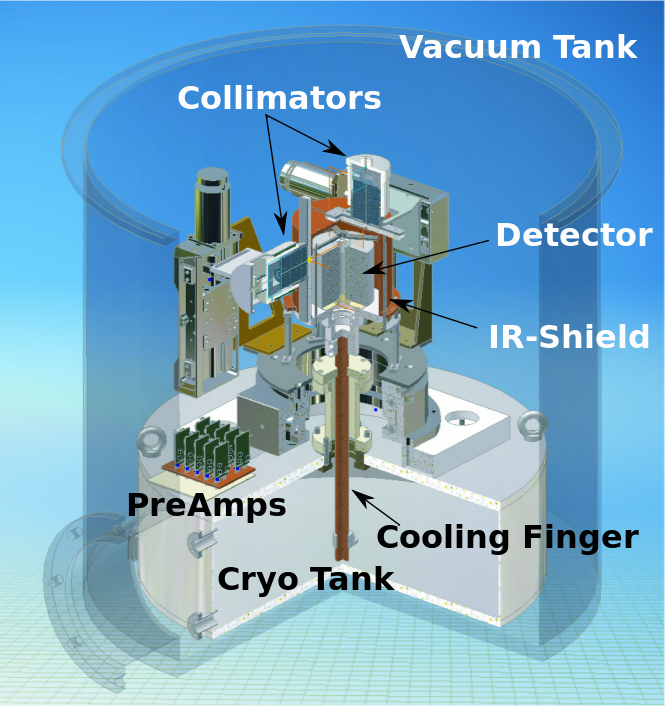}
  \end{center}
  \caption{Schematic view of the interior of the GALATEA test-facility.} 
  \label{fig:setup}
\end{figure}

The GALATEA test-facility is based on a vacuum tank housing a system of three precision motorized stages surrounding the detector holder \cite{Irlbeck}. The latter is cooled by a liquid-nitrogen tank (LN-tank) via a copper cooling finger. The three-stage\cite{pimicos} system is used to move two collimated radioactive sources, one across the top and one around the mantle of the detector. The collimators are moved along two slits of an infrared (IR) shield which surrounds the detector (see section \ref{ir}). This system allows an almost complete surface scan of a cylindrical detector. Figure~\ref{fig:setup} depicts the schematic of GALATEA.

\subsection{The vacuum chamber}
\label{tank}
\begin{figure}[!ht]
 \begin{minipage}[c]{.5\textwidth}
 \centering
 \includegraphics[width=1\textwidth]{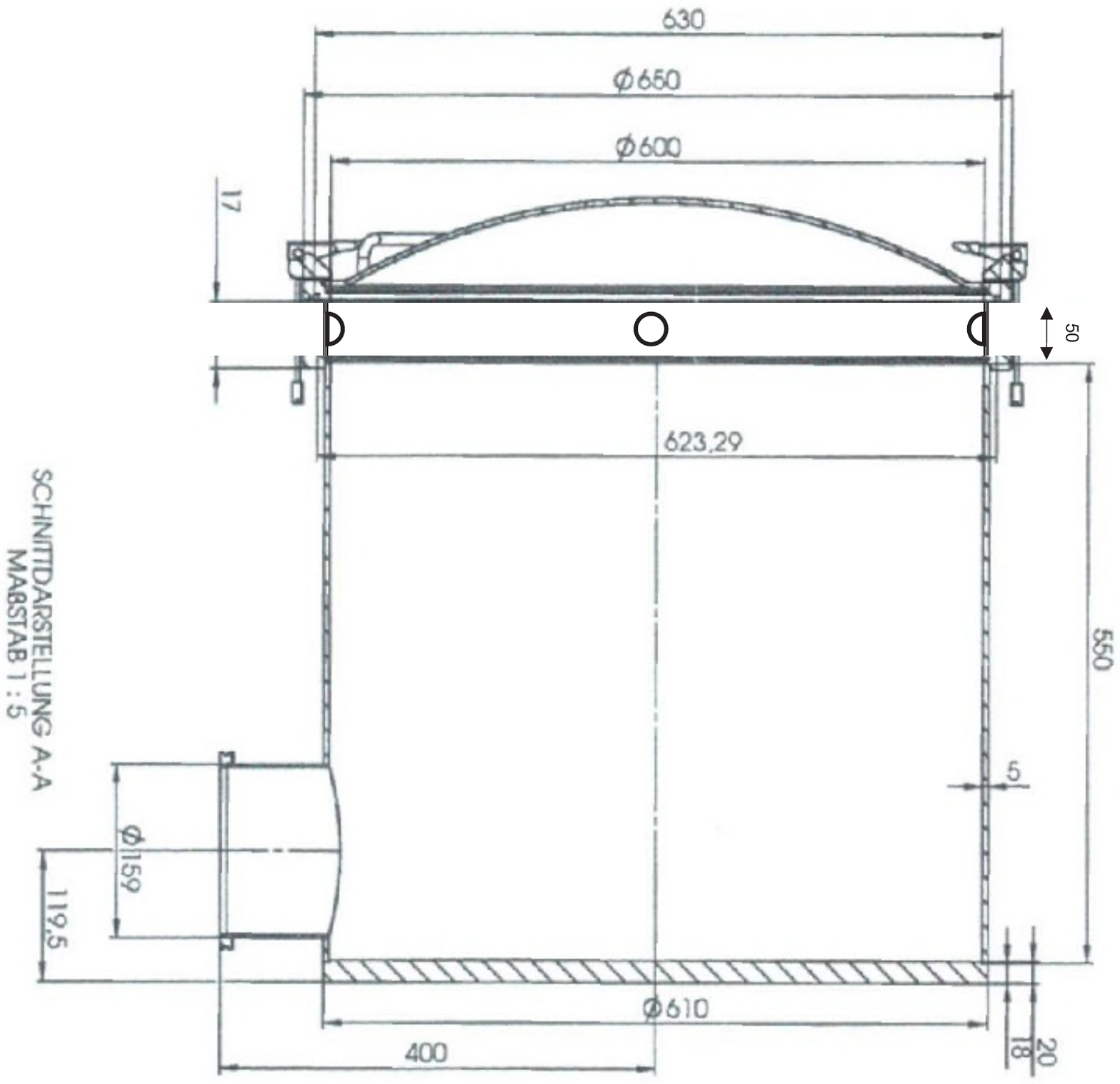}
 \subcaption{}
\end{minipage}
\begin{minipage}[c]{.5\textwidth}
 \centering
 \includegraphics[width=1\textwidth]{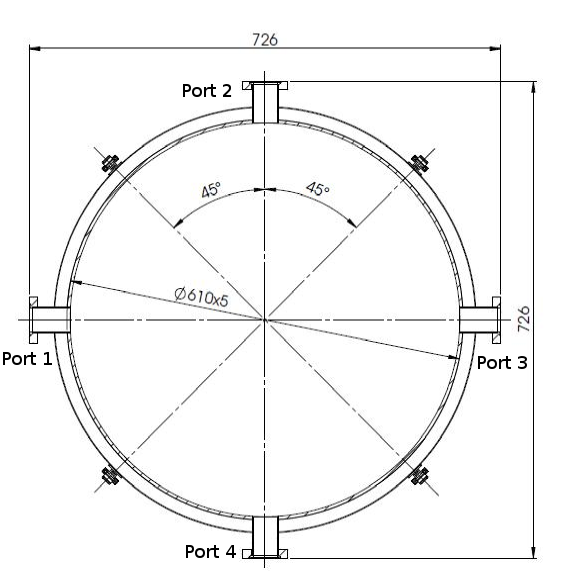}
 \subcaption{}
\end{minipage}
\caption{a) Technical drawing of the main vacuum chamber and the lid, b) technical drawing \cite{Irlbeck} of the modular chamber housing the measurement ports.} 
\label{fig:chamber}
\end{figure}

The main vacuum chamber, which basically contains the whole setup, is based on a 65\,cm high stainless steel cylinder with an inner diameter of 60\,cm and a wall thickness of 5 mm, see Fig.~\ref{fig:chamber} (a). The system is closed by a stainless-steel lid.
Between lid and body, a modular chamber with four DN 40 CF measurement ports is mounted, see Fig.~\ref{fig:chamber} (b). Two of the ports host two different pressure gauges, a PKR 251, Active Pirani/Cold Cathode Transmitter \cite{pirani} (single gauge) and a Bayard-Alpert-Type \cite{barion} (Barion) sensor. A third port hosts an inlet vent to purge the system with gaseous nitrogen. The presence of two pressure gauges with different operating ranges (single gauge: from $\approx 10^{-9}$\,mbar to 1\,bar, Barion: from $\approx 10^{-11}$\,mbar to $10^{-2}$\,mbar) in different positions assures an excellent monitoring of the vacuum quality.
Towards the bottom of the chamber, there is a DN 160 ISO-K weld-on nozzle which connects the main volume to a DN 160 ISO-K stainless steel 3D cross with six DN 160 ISO-K flanges \cite{vacom}. The cross hosts the cable feed-throughs serving the internal instrumentation, the motors of the stages, the detector read out and pipes for the LN-tank. A turbo pump (TMH 521 P, Pfeiffer Vacuum, 833 rotations/s pumping speed) is connected to one of the DN 160 ISO-K flanges of the cross through a VAT gate-valve (shutter). The shutter is used to disconnect the vacuum volume from the pump in order to eliminate microphonic effects which, otherwise, would make the detector operation impossible. 
 
All seals, but the one for the lid, are metallic. The lid is sealed with a viton O-ring. The viton O-ring preserves a good vacuum while allowing to open and close the facility multiple times. The use of metallic seals in all the other flanges makes it easier to maintain a good vacuum ($\mathcal{O}(10^{-6})$\,mbar) for several days without pumping.

\subsection{The cryogenic system}
\label{cryo}
\begin{figure}[!ht]
  \begin{center}
    \includegraphics[scale=0.7]{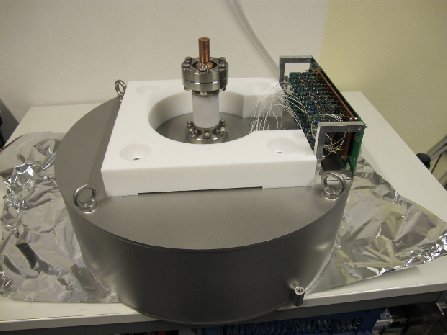}
  \end{center}
  \caption{On top of the cylindrical cryogenic tank, a PTFE plate (white) surrounds the ceramic flange for the cooling finger. On the right side, an aluminum structure, housing the readout electronics, is mounted.} 
  \label{fig:cryopic}
\end{figure}

Germanium detectors have to be operated ideally at about 100\,K. GALATEA is equipped with an internal 30\,l stainless steel LN-tank. It is located on the bottom of the main vacuum chamber and has a cylindrical shape with an outer diameter of 48\,cm and a total height of about 16 cm, see Fig.~\ref{fig:cryopic}. The outer surface of the LN-tank acts as an intrinsic cryopump contributing to the good vacuum performances of the system. Additional cryopumping zeolith is not necessary.

The LN-tank rests on three bolts with a contact area of $\approx 7$ mm$^2$ each to minimize its thermal coupling to the outside wall.
A copper cooling-finger, 290\,mm long, 16\,mm in diameter, emerges through a ceramic flange on top of the LN-tank. Ceramic was chosen to electrically decouple the detector holder, which is  sitting on top of the cooling-finger, from the rest of the setup. 
As shown in Fig.~\ref{fig:cryopic}, the top of the LN-tank is covered with a Polytetrafluoroethylene (PTFE) plate, 27\,x\,27\,x\,4.5\,cm, supporting the three-stage system. It acts both as a thermal and an electrical insulator between the motorized stages and the LN-tank. 
 
To thermally insulate the inside from the outer wall, a thermal shield, made of 3 layers of super-insulation foil, COOLCAT\footnote{Multi-layer cryogenic foil: each layer consists of 10 spot welded sub-layers, made of double-sided aluminized 6 $\mu m$ polyester-film (perforated), interleaved with 10 sub layers of polyester knit-woven spacer \cite{coolcat}.}, covers most of the inner surface. A mesh (stainless steel) is used to hold the insulation foil approximately 1\,cm apart from the outer wall. Furthermore, the sides of the LN-tank itself are also covered by three layers of COOLCAT.  

\subsection{The infrared shield and the detector holder}
\label{ir}
\begin{figure}[!ht]
 \begin{minipage}[c]{.5\textwidth}
 \centering
 \includegraphics[width=1\textwidth]{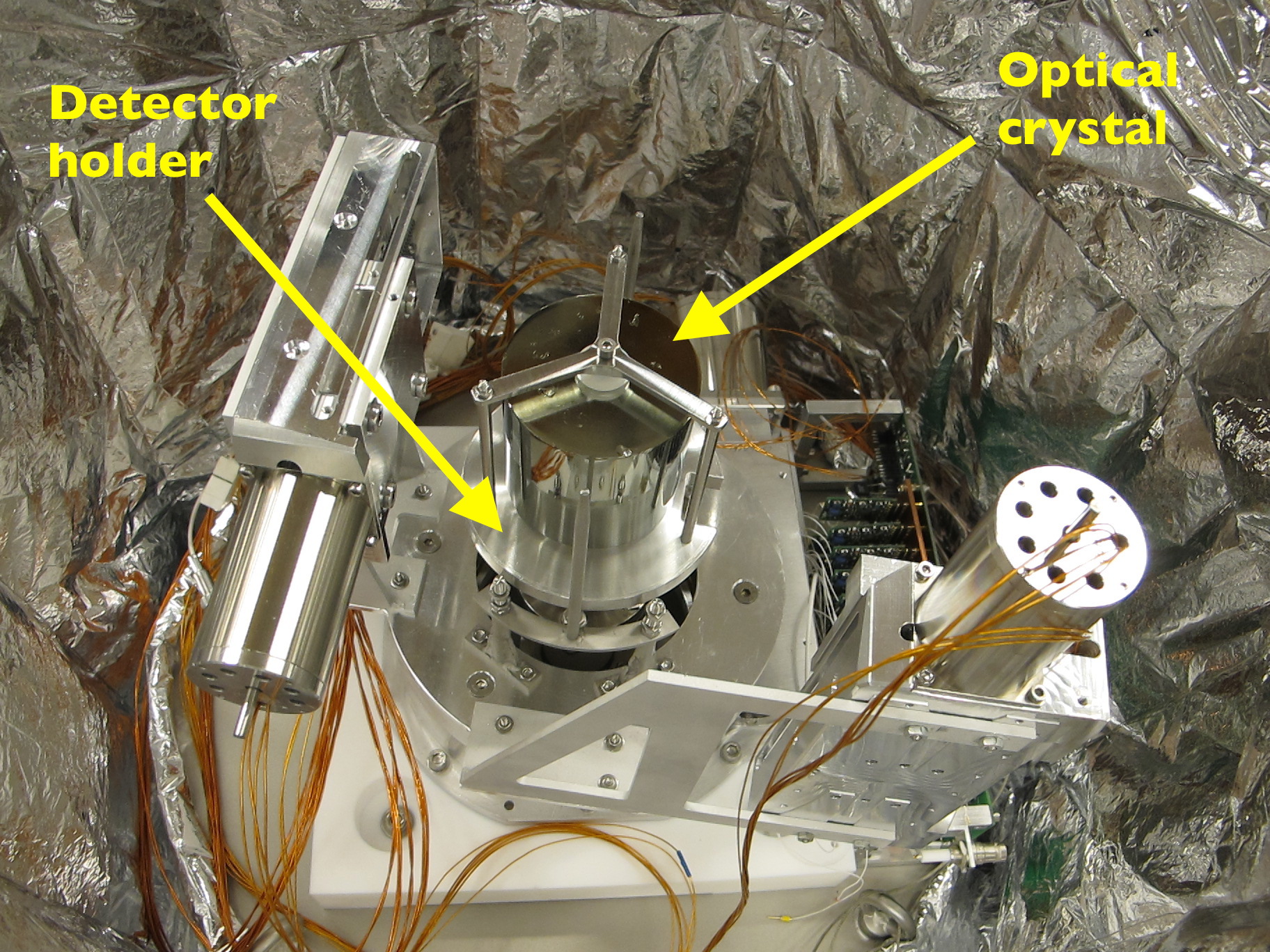}
 \subcaption{}
\end{minipage}
\begin{minipage}[c]{.5\textwidth}
 \centering
 \includegraphics[width=1\textwidth]{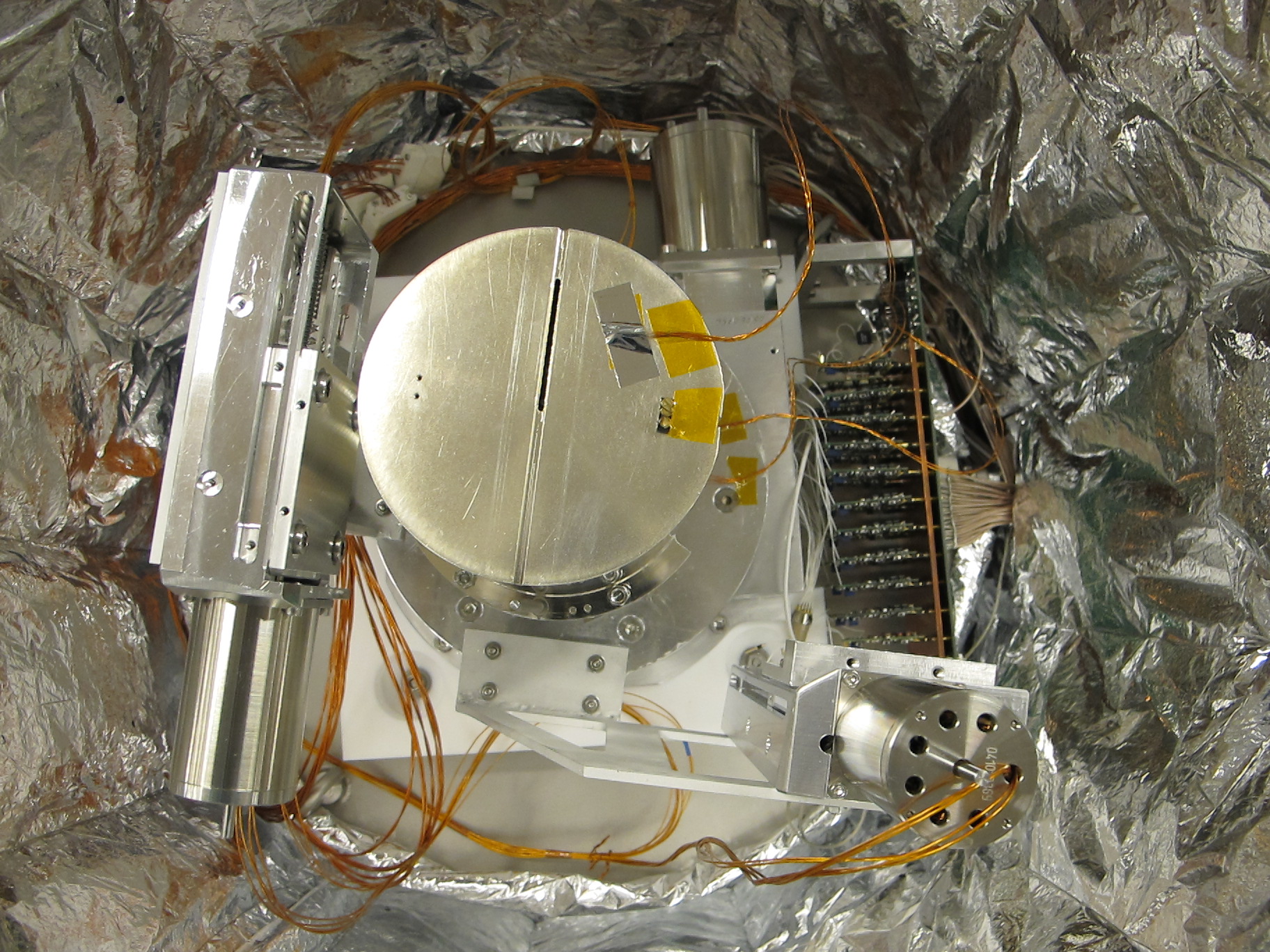}
 \subcaption{}
\end{minipage}
\caption{a) Detector holder installed in GALATEA with an optical crystal used for thermal tests, b) the IR shield placed around the detector; the top slit is clearly visible.} 
\label{fig:holder}
\end{figure}

Infrared radiation on the detector will increase the detector temperature and, thus, the leakage current. Therefore, the detector is surrounded by a cylindrical IR shield made out of copper. It was electropolished and silver coated to both minimize oxidation and achieve good reflectivity for IR radiation. It is 109\,mm high and has an inner diameter of 110 mm. The IR shield is in direct thermal contact with the base-plate of the cooled detector holder, see Fig.~\ref{fig:holder}. 
The IR shield has two slits, about 2\,mm wide, one on the top and one on the side. The collimators slide along these slits, moved by two linear precision stages. The shield and the two linear stages are rotated by a circular stage, see Sec. \ref{motor}. To avoid grounding problems, the IR shield and thus the detector are carefully electrically decoupled from the motor system by means of threaded PTFE tubes and nuts. The PTFE also provides a good thermal insulation.

\subsection{The three-stage system and the collimators}
\label{motor}
\begin{figure}[!ht]
  \begin{center}
    \includegraphics[scale=0.5]{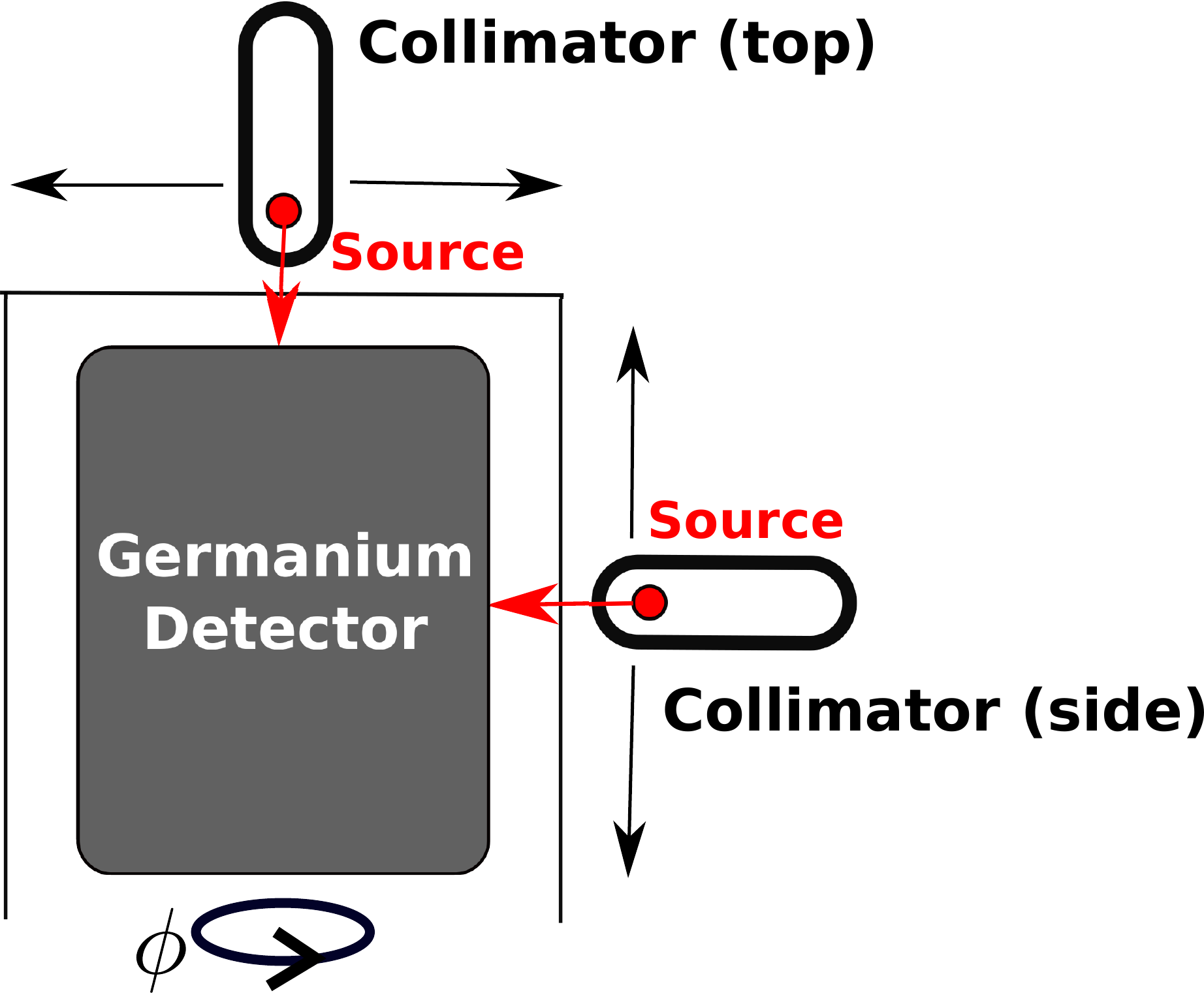}
  \end{center}
  \caption{Two linear motors (PLS-85 UHV CRYO, PImiCos), one on top and one on the side of the detector, sit on a rotation stage (DT-120 UHV CRYO, PImiCos). Figure taken from \cite{Irlbeck}.}
  \label{fig:scan}
\end{figure}

Three ultrahigh vacuum (UHV) motorized stages, designed for cryogenic temperatures between 80 and 100 K, move two collimated sources. Figure~\ref{fig:scan} shows a sketch, illustrating the scanning principle. The linear stages have a precision of $\approx \, 1 \,\mu m$ while the rotational one has a precision of about 0.02$^{\circ}$.

For commissioning, a collimated $^{241}$Am $\alpha$-source was used on top of the detector and a collimated $^{152}$Eu gamma source was mounted on the side. The contact material between the sliding collimators and the IR shield is black Murtfeldt \cite{murtfeldt}, a plastic causing almost no friction between the two components, suitable for cryogeninc temperatures and not transparent to IR radiation.
 
\begin{figure}[!ht]
 \begin{minipage}[c]{.5\textwidth}
 \centering
 \includegraphics[width=1\textwidth]{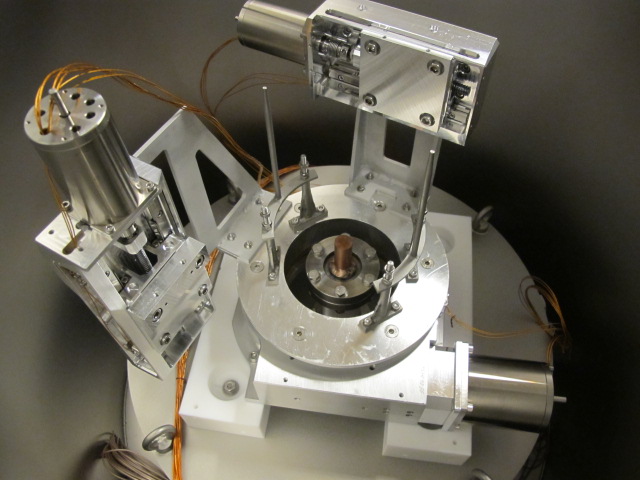}
 \subcaption{}
\end{minipage}
\begin{minipage}[c]{.5\textwidth}
 \centering
 \includegraphics[width=1\textwidth]{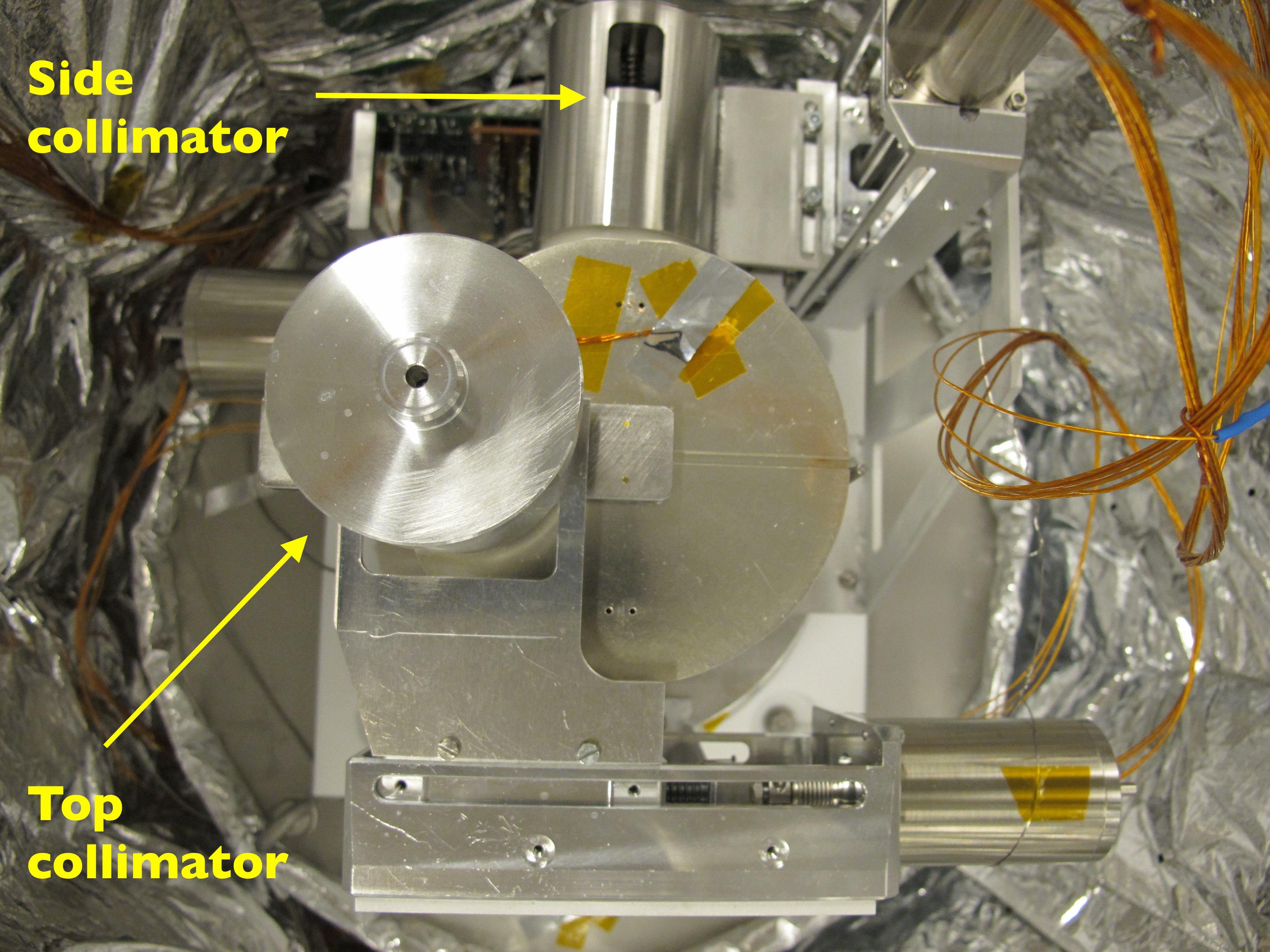}
 \subcaption{}
\end{minipage}
\caption{a) The three stage system, b) the stage plus collimator system. The top slider has a range of about 52 mm, the vertical one covers about 76\,mm along the side of the detector while the rotation stage can go from 0 to $\approx$\,350 degrees. } 
\label{fig:stage}
\end{figure}

The aluminum collimator supports are fixed to their respective stages. These cylindrical supports are 60\,mm high with an outer diameter of 58\,mm and an inner diameter of 47 mm. Each support hosts a cylindrical source holder plus 5 cylindrical tungsten segments, stacked behind each other. Each tungsten segment is 10\,mm thick and has a central bore hole. During commissioning, the $^{241}$Am source had three segments placed between the source and the detector, one with a 3.2\,mm diameter hole and two with 3\,mm diameter hole. This configuration results in a beam spot on the detector top surface of 4.4\,mm diameter. The $^{152}$Eu source had three segments with 3\,mm diameter hole in front of the detector, providing a beam spot on the mantle of the target detector of of 3.6\,mm diameter. The use of tungsten collimator segments provides a basically perfect collimation of the low-energy gamma lines of the $^{152}Eu$ source.
Figure~\ref{fig:stage} shows the stage plus collimator system.

\subsection{Supersiegfried, a special detector prototype}
\label{susie}
\begin{figure}[!ht]
 \begin{minipage}[c]{.4\textwidth}
 \centering
 \includegraphics[width=1\textwidth]{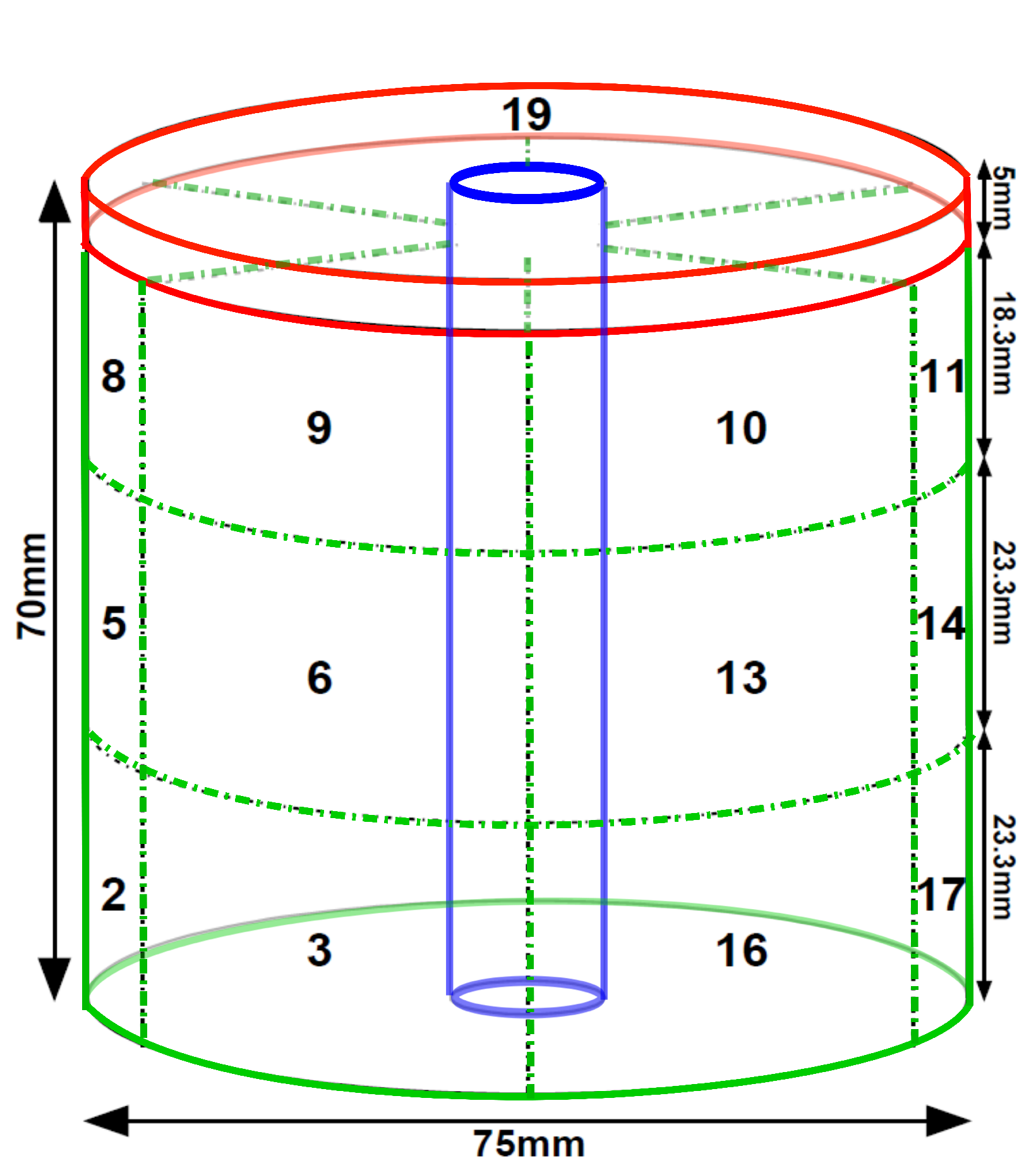}
 \subcaption{}
\end{minipage}
\hskip 1cm
\begin{minipage}[c]{.5\textwidth}
 \centering
 \includegraphics[width=1\textwidth]{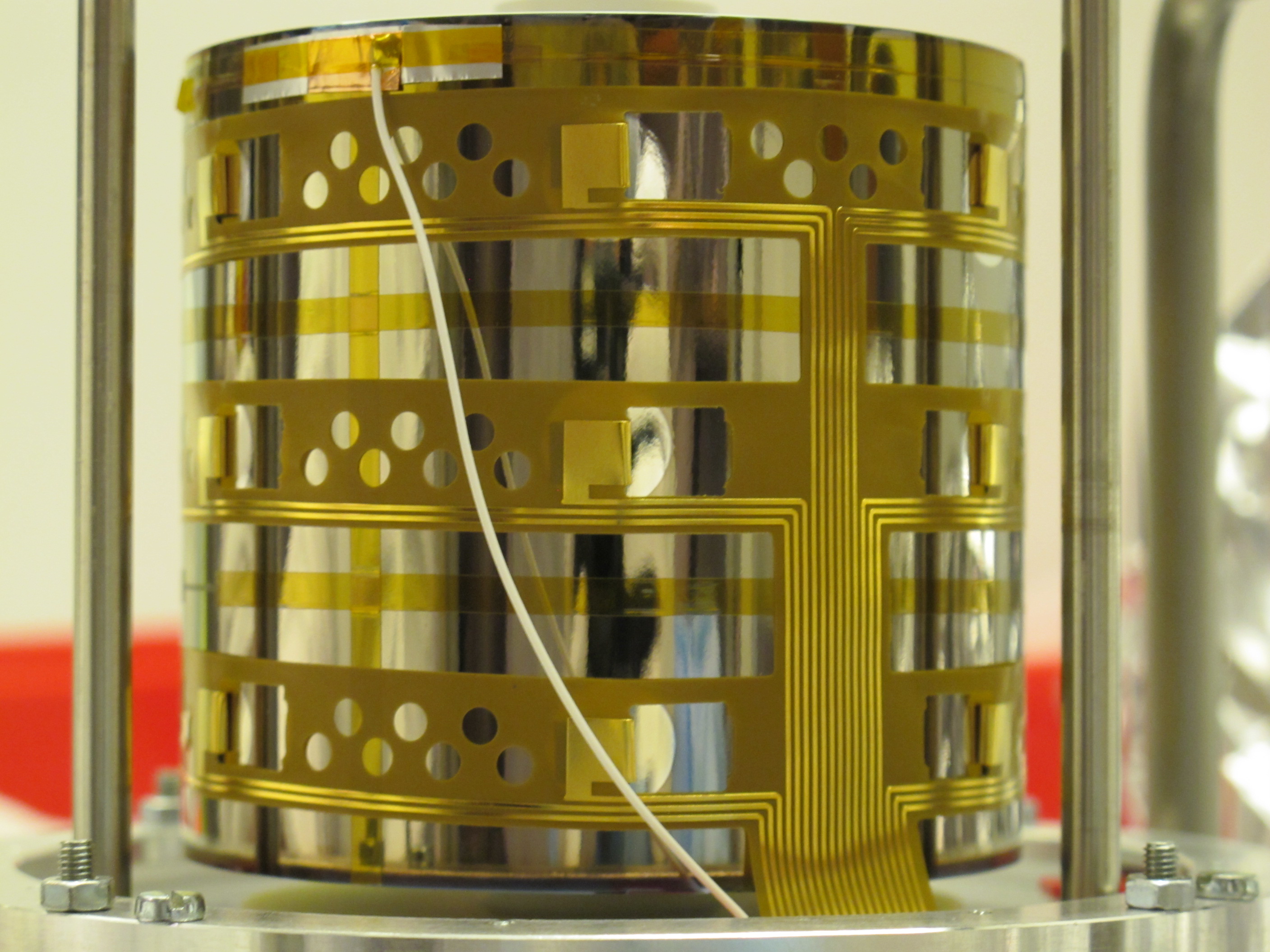}
 \subcaption{}
\end{minipage}
\caption{a) Schematic of the 19-fold segmented detector, b) picture of SuSie.} 
\label{fig:susie}%left--> 5.3 sabine, right --> picture of my ADP presentation
\end{figure}

A special true-coaxial n-type HPGe detector, manufactured by CANBERRA \cite{canberra}, was mounted and tested in GALATEA during commissioning. This detector, called Supersiegfried (SuSie), is 70\,mm high, has an inner bore hole radius of 5.05\,mm and an outer radius of 37.5\,mm, and is operated with a bias voltage of +3000\,V. The detector mass is 1634.5\,g.
It has a 5\,mm thick segment on top of an 18-fold segmentation \cite{18fold}. The thin extra layer, 19$^{th}$ segment, on top\footnote{A cylindrical coordinate system is used with the origin at the center of the detector and Z pointing upwards; $\phi$ denotes the azimuthal angle.} is unsegmented in $\phi$.
It was especially designed to study surface channel effects and charge trapping \cite{surface}.  
See Fig.~\ref{fig:susie} for schematic and dimensions of the segments.

\subsection{Electronics and data acquisition}
\label{elec}
\begin{figure}[!t]
  \begin{center}
    \includegraphics[scale=0.6]{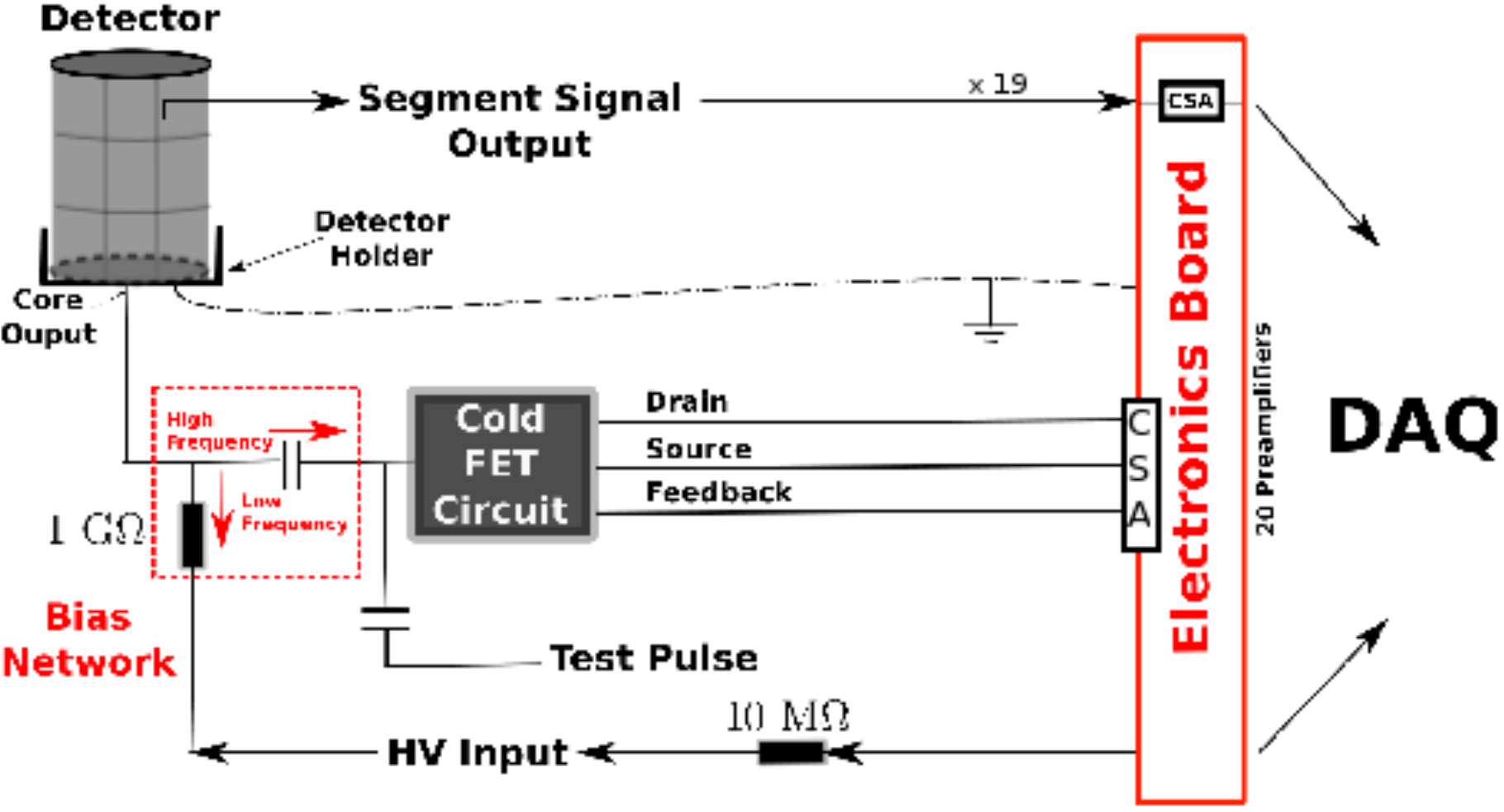}
  \end{center}
  \caption{Schematic of the electronics of GALATEA, taken from \cite{Irlbeck}.}
  \label{fig:elec}
\end{figure}

As shown in Fig.~\ref{fig:cryopic}, an aluminum structure on top of the LN-tank houses the main electronics board. It is mounted as close as possible to the detector in order to minimize the length\footnote{Each cm of cable adds 1\,pF to the input capacitance affecting the preamplifier.} of the read-out cables. The board supports 20 charge sensitive preamplifiers (CSA), PSC-823V produced by CANBERRA \cite{canberra}, working with a RC-feedback circuit. The preamplifiers service the core and all the segment signals of the detector. The Junction Field Effect Transistor (JFET), the first pre-amplifier stage, is integrated on the board for the segment circuits. For the core circuit, the core JFET is integrated in the detector holder, right below the detector. The grounding of the detector is provided from the preamplifier board via a massive copper band tightened to the cooling-finger on which the detector is mounted. 
A standard high voltage NHQ 206L iseg \cite{iseg}, HV, power supply is used to bias the detector. 
Figure~\ref{fig:elec} shows a schematic of the electronics of the GALATEA teststand. More details can be found elsewhere \cite{Irlbeck}.
    
The digitization of the preamplified signals takes place outside the vacuum chamber. A digital multichannel data acquisition (DAQ), XIA PXI Compact PCI, model PXHI-18 \cite{pixie}, with 75 MHz sampling rate is used. The DAQ records not only the energy and the time of the event but also the pulse in an up to 15\,$\mu s$ long time window. This allows for offline pulse shape analysis. An offline energy reconstruction is therefore possible. A flat-top software-filter is used for the online energy reconstruction.

\newpage
\section{The online-monitoring and the automatic LN$_2$-refilling system}
\label{monitor}
\begin{figure}[!ht]
  \begin{center}
    \includegraphics[scale=0.2]{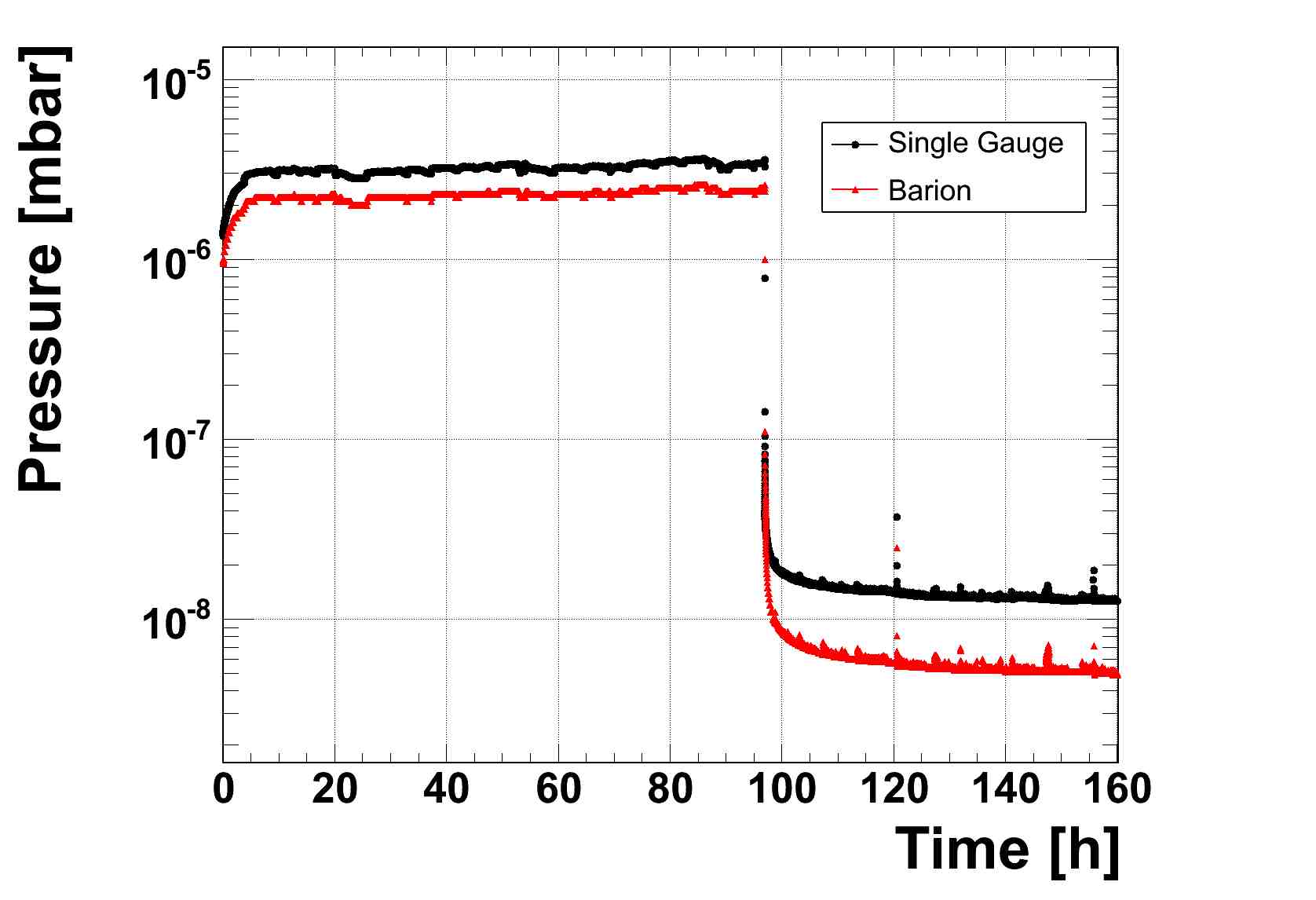}
  \end{center}
  \caption{Pressure during measuring (higher pressure) and pumping (lower pressure) phases. The circles refer to a full range single gauge sensor, the triangles refer to a barion sensor.}
  \label{fig:pressure}
\end{figure}

The GALATEA operational conditions are monitored by a multiple-sensor system. The pressure, the temperature of different components, and the fill-level of the LN-tank are constantly recorded.
As mentioned in Sec. \ref{tank}, the setup is equipped with two pressure sensors. Figure~\ref{fig:pressure} shows the pressures as measured by the two sensors during operation. Two different regimes are clearly visible. The lower pressure, $\mathcal{O}( 10^{-8} )$\,mbar, regime corresponds to the pumping, i.e. non-measuring, phase. The higher pressure corresponds to the measuring phase when the turbo pump is turned off and its volume is separated from the vacuum volume after closing the shutter. As can be seen, the GALATEA tank ensures a vacuum of $\mathcal{O}( 10^{-6} )$\,mbar for more than 100 hours without pumping, allowing for extended measuring phases. This vacuum performance was achieved by rigorous conditioning during which the chamber and all other inside parts, but the detector and the electronics board, were heated up to $\approx 150^{\circ}$C for several weeks while pumping. The electronics board together with the preamplifiers were conditioned separately at a temperature of about $80^{\circ}$C. Suitable components were previously cleaned in an ultrasonic bath and wiped with isopropanol. 
 
\begin{figure}[!ht]
  \begin{center}
    \includegraphics[scale=0.25]{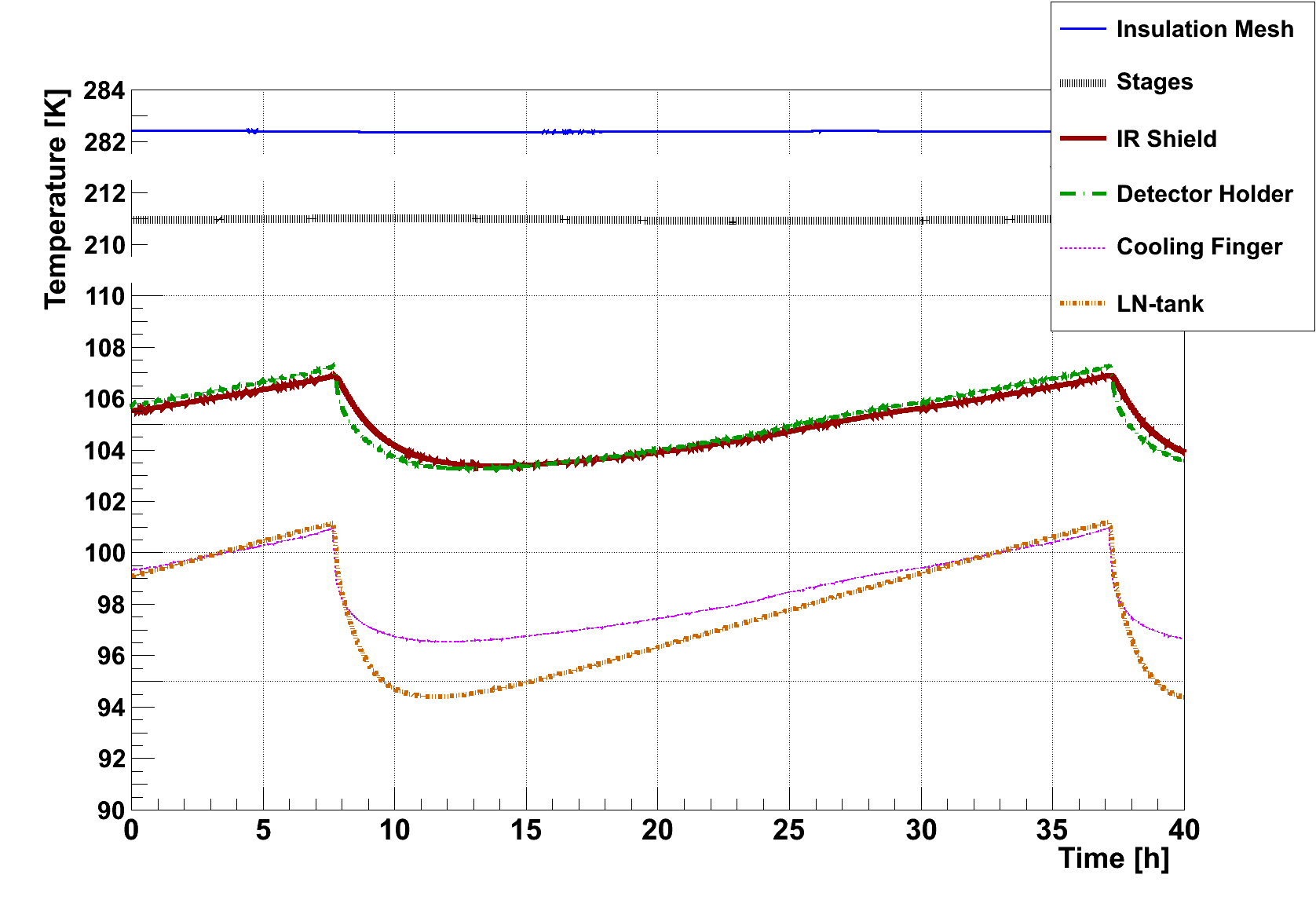}
  \end{center}
  \caption{Temperature of different components inside GALATEA.}
  \label{fig:temperature}
\end{figure}

The temperature is monitored using PT100 \cite{pt100} temperature sensors placed in six different places inside the vacuum volume. The sensors are connected to the LN-tank, the cooling-finger, the detector holder, the IR shield, the stages and the mesh holding the COOLCAT insulation foil. These sensors are fixed to the components with a vacuum compatible glue, NEE-001 from Dr. Neumann Peltier Technik \cite{nee}, in order to guarantee a good and uniform thermal contact.
In Fig.~\ref{fig:temperature}, the temperature behavior is shown. As expected, the coldest parts are the LN-tank, the cooling-finger, the detector and the IR shield. As these components have the best thermal contact to the LN$_2$, their temperatures are the most sensitive to the fill-level inside the LN-tank. This is reflected in the oscillating pattern of their temperatures, as visible in Fig.~\ref{fig:temperature}. The oscillation amplitude for the detector holder is $\approx$ 4\,K. The knowledge of the temperature changes of the detector, is very important since these influence the pulse length \cite{pulselength}. 

\begin{figure}[!ht]
  \begin{center}
    \includegraphics[scale=0.5]{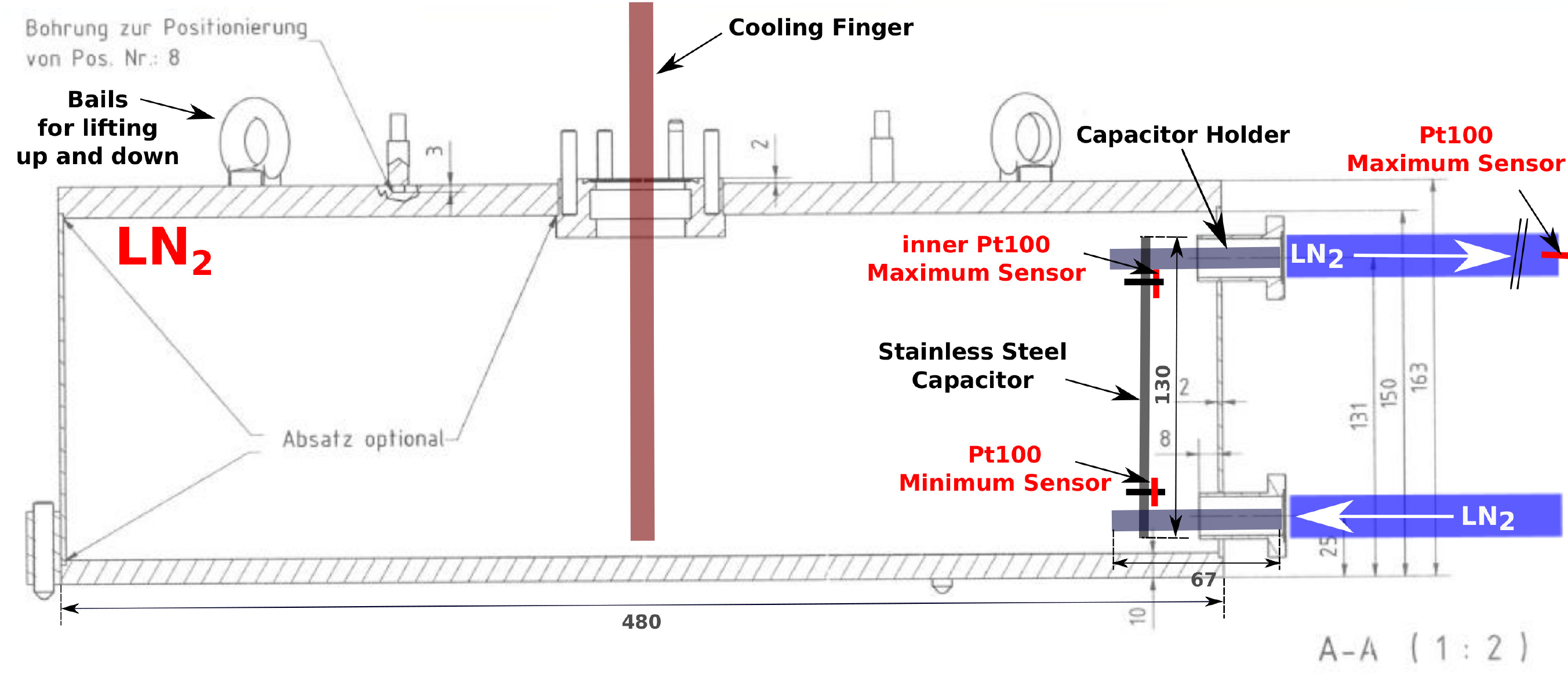}
  \end{center}
  \caption{Technical drawing of the LN-tank together with the LN$_2$ monitoring and refilling sensors system \cite{Irlbeck}.}
  \label{fig:ln2monitor}
\end{figure}

The fill-level of the LN-tank is monitored via a quadrupole capacitance measurement of a stainless-steel double cylinder capacitor \cite{cobra}, \cite{schulz}. The working principle is to measure the capacitance of a double cylinder capacitor which is located inside the LN-tank, submerged in the LN$_2$. The quadrupole measurement is performed by using two PT100 sensors, one placed at the bottom of the capacitor (minimum sensor) and one at the top (inner maximum sensor), as shown in Fig.~\ref{fig:ln2monitor}.

\begin{figure}[!ht]
  \begin{center}
    \includegraphics[scale=0.7]{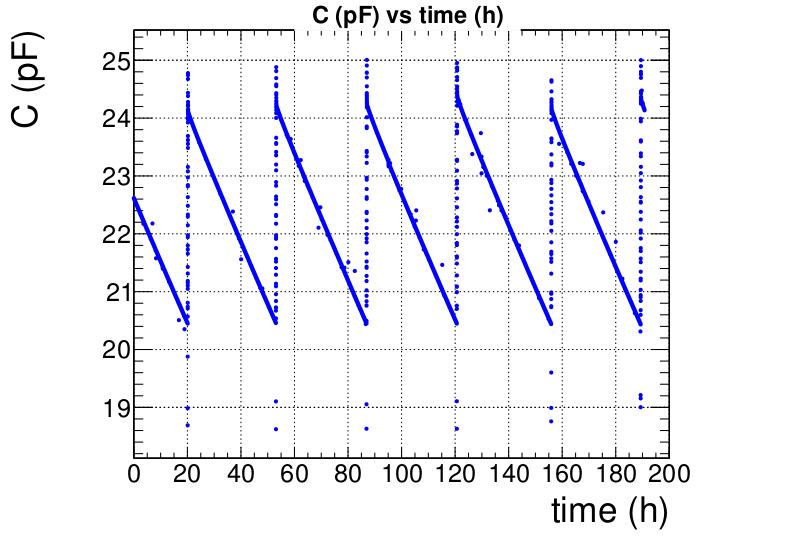}
  \end{center}
  \caption{Time evolution of the capacitance of the double-cylinder capacitor monitoring the LN$_2$ fill level. This value is proportional to the LN$_2$ fill level.}
  \label{fig:C}
\end{figure}

There is a linear relation between the capacitance (C) of the double capacitor and the LN$_2$ fill level inside the capacitor, i.e. inside the LN-tank.
This linearity is clearly demonstrated in Fig.~\ref{fig:C}. When the Ln-tank is full C\,$\approx$\,24\,pF, while C\,$\approx$\,18.5\,pF for an empty LN-tank. An automatic refill from an external LN-reservoir is started when C\,$\approx$\,20.5\,pF. The good thermal insulation of the system allows a LN$_2$ refilling cycle of about 40 hours, which causes the temperature oscillations of Fig.~\ref{fig:temperature}.
 
\section{Commissioning}
\label{results}

\begin{figure}[!ht]
  \begin{center}
    \includegraphics[scale=0.25]{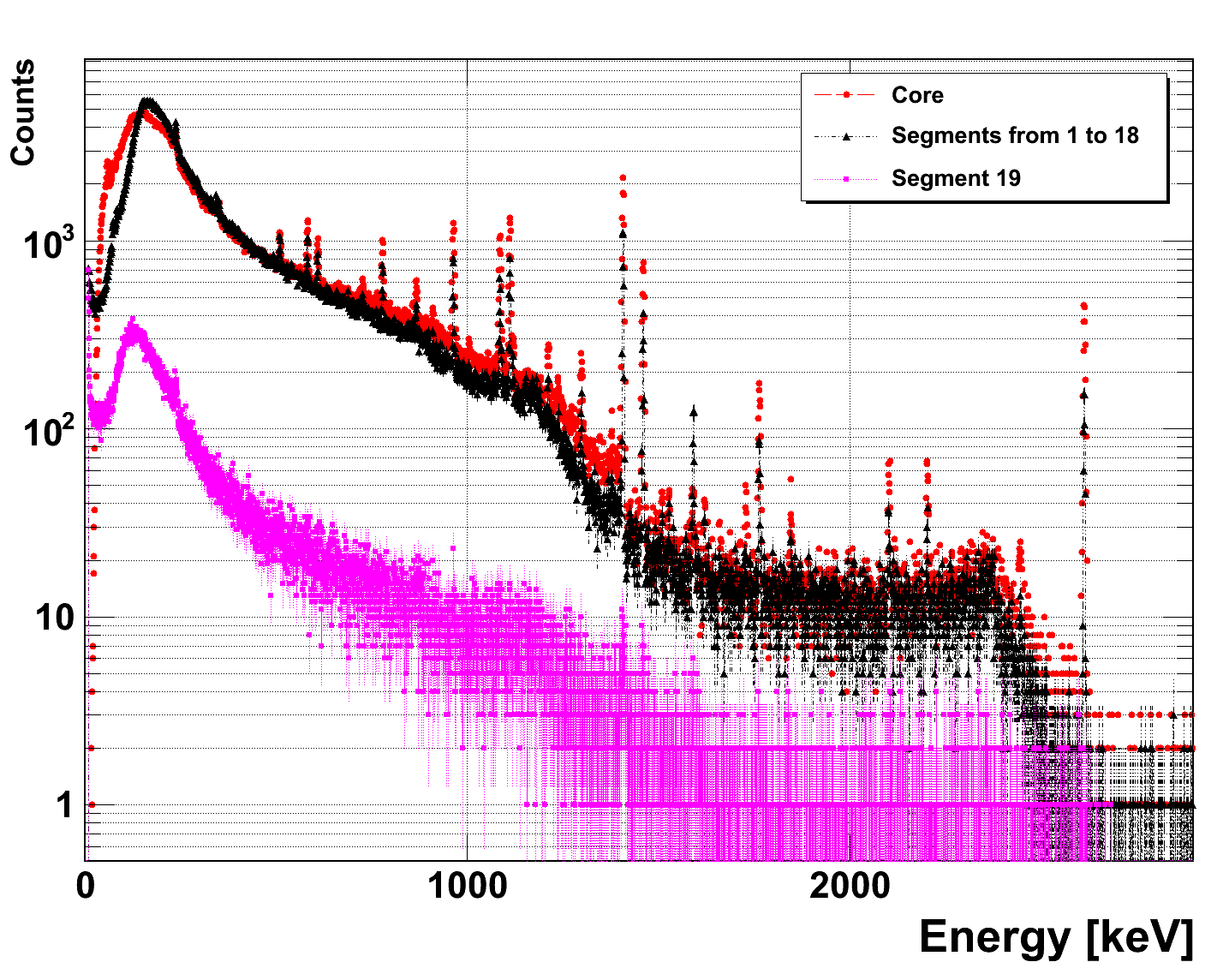}
  \end{center}
  \caption{Calibrated and crosstalk corrected $^{228}$Th spectra for the core (circles), the sum of the spectra of segments 1 to 18 (triangles) and the 19th segment (squares). }
  \label{fig:spectra}
\end{figure}

Figure~\ref{fig:spectra} shows the energy spectra of the core, the 19th segment and the sum of the spectra of segments 1 to 18. They were taken during a calibration run using an uncollimated $^{228}$Th source placed outside the vacuum tank. The $^{241}$Am and $^{152}$Eu sources were removed as far as possible from the detector.  After crosstalk correction, the spectra were calibrated using the most prominent gamma lines from the $^{228}$Th spectrum. The energy shown here represents the online DAQ energy reconstruction. 
More details about the crosstalk correction and calibration procedures can be found elsewhere \cite{Irlbeck}.   

The energy resolution, defined as the FWHM of a gamma-peak, was evaluated for some of the most prominent peaks for both the core and the sum of the segments (1 to 18), see Table \ref{tab:resolution}. Each peak was fitted with a Gaussian plus a first order polynomial function. 
The energy resolutions of the core and the 18 segments at 2.6\,MeV  are (5.92 $\pm$ 0.04)\,keV and (3.52 $\pm$ 0.05)\,keV respectively. 

\begin{table}[!ht]
\begin{center}
\begin{tabular}{c|c|c}
\hline
          & \textbf{Energy [keV]} & \textbf{FWHM [keV]}\\
\hline
\hline
  & 511 & 5.05 $\pm$ 0.17\\
  & 728 & 4.19 $\pm$ 0.32\\
\textbf{Core} & 1408 & 5.44 $\pm$ 0.02 \\
  & 1460 & 5.71 $\pm$ 0.05 \\
  & 2204 & 5.52 $\pm$ 0.25 \\
  & 2614 & 5.92 $\pm$ 0.04 \\
\hline
\hline
   & 511 & 3.14 $\pm$ 0.14\\
  & 728 & 2.63 $\pm$ 0.21\\
\textbf{Segments} & 1408 & 3.34 $\pm$ 0.02 \\
  & 1460 & 3.10 $\pm$ 0.05 \\
  & 2204 & 4.23 $\pm$ 0.32 \\
  & 2614 & 3.52 $\pm$ 0.05 
\end{tabular}
\end{center}
\caption{Energy resolution \cite{Irlbeck} for different gamma peaks, for both the core and the overlay of segments 1 to 18. }
\label{tab:resolution}
\end{table}

\begin{figure}[h!]
  \begin{center}
    \includegraphics[scale=0.25]{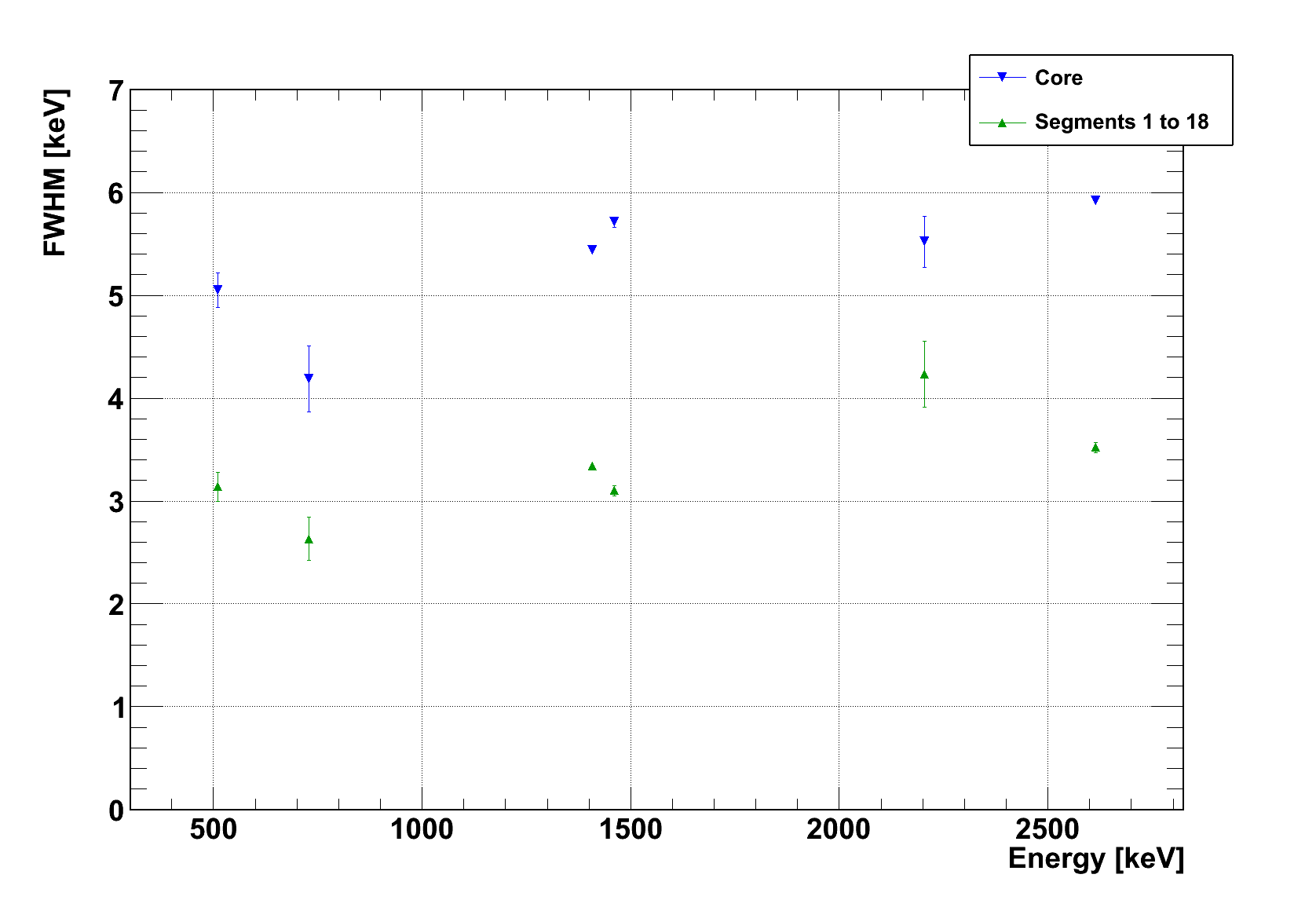}
  \end{center}
  \caption{FWHM vs energy, both for the core (downward triangles) and for the sum of segments 1 to 18 (upward triangles).}
  \label{fig:resolution}
\end{figure}

A graphical representation of the values of Table~\ref{tab:resolution} is shown in Fig.~\ref{fig:resolution}. The intrinsic FWHM is expected to be proportional to the energy \cite{owens}. However, no significant dependence is observed. This  indicates that the dominant contribution to the energy resolution comes from the electronics. 

%alpha-data
As previously stated, a $^{241}$Am source was placed inside the top collimator to irradiate the passivated top surface of SuSie with alpha particles.
  \begin{figure}[!ht]
  \begin{center}
    \includegraphics[scale=0.4]{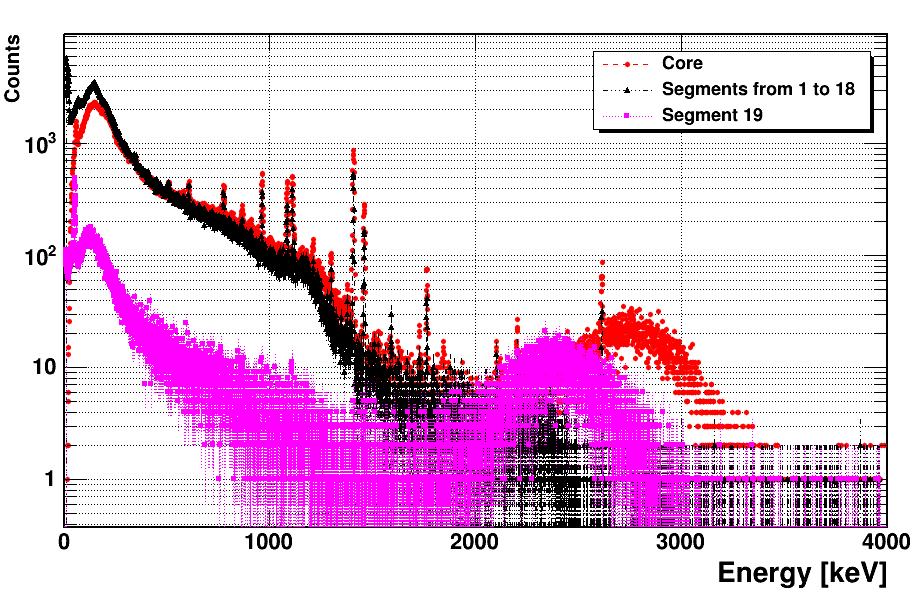}
  \end{center}
  \caption{Energy spectra as measured with the core (circles), the segments 1 to 18 summed up (triangles) and the 19th segment (squares). \cite{garbini}}
  \label{Fig:alphaSpectrum}
\end{figure}
Typical energy spectra obtained with this source are shown in Fig.~\ref{Fig:alphaSpectrum}. Broad bumps due to the alpha particles are visible between 2 and 3\,MeV for the core and segment 19. This feature does not arise in the energy spectra of segments 1 to 18 due to the minimal penetration depths of alpha particles. An alpha particle of $\approx$5\,MeV, in germanium,  only penetrates around 20\,$\mu$m. Therefore the observation of alpha-interaction demonstrates that the dead layers are extremely thin in the region under investigation.
The energy recorded corresponds to the energy deposition of the alpha particles in the active volume of the detector after passing through a region with reduced charge collection efficiency (dead layer). Thus, the mean energy of the alpha-bump depends on the thickness of the effective dead layer. The presence of a dead layer is due to the passivation layer together with a distorted electric field right underneath. The fact that the alpha-induced bumps are located at different energies for core and segment 19 indicates that electrons and holes are trapped with different probabilities.
Events in segment 19 with energies compatible to $\alpha$-interactions do not show energy deposition in any of the other 18 segments.
 A more detailed and quantitative analysis will be published separately.

\clearpage

\section{Summary and Outlook}
\label{out}
The GALATEA test-facility is a novel device, especially designed for detailed studies of germanium detectors. It is now fully operational. Studies with a collimated $\alpha$-source are ongoing. A first analysis reveals that electrons and holes are trapped with different probabilities close to the passivated surfaces of a germanium detector. Further studies with low-energy gamma- and $\beta$-radiation are planned. The installation of a tunable infrared laser is foreseen for more detailed studies of surface effects. Changing the wavelength of the infrared laser beam changes its penetration depth. This will allow to probe the surface at different depths with high precision. This kind of study can improve the understanding of germanium detectors, especially in the proximity of the electric field distortions.

\section{Acknowledgments}
\label{ack}
We would like to thank the technical department of the Max-Planck-Institut f\"{u}r Physik for the strong support.

\bibliographystyle{elsarticle-num}
\bibliography{<your-bib-database>}

\end{document}